\documentstyle[psfig,times]{mn}

\newif\ifAMStwofonts

\def\etal{et al.\ }

\def\approxlt{\mathrel{\spose{\lower 3pt\hbox{$\sim$}}
	\raise 2.0pt\hbox{$<$}}}
\def\approxgt{\mathrel{\spose{\lower 3pt\hbox{$\sim$}}
	\raise 2.0pt\hbox{$>$}}}
	
\def\cm{{\rm\thinspace cm}}

\def\deg{$^\circ$}
\def\erg{{\rm\thinspace erg}}
\def\eV{{\rm\thinspace eV}}

\def\keV{{\rm\thinspace keV}}
\def\m{{\rm\thinspace m}}

\def\km{{\rm\thinspace km}}
\def\kpc{{\rm\thinspace kpc}}
\def\Lsun{\hbox{$\rm\thinspace L_{\odot}$}}
\def\rad{{\rm\thinspace rad}}

\def\Mpc{{\rm\thinspace Mpc}}
\def\Msun{\hbox{$\rm\thinspace M_{\odot}$}}

\def\s{{\rm\thinspace s}}

\def\ergps{\hbox{$\erg\s^{-1}\,$}}

\def\kmps{\hbox{$\km\s^{-1}\,$}}

\def\pcmsq{\hbox{$\cm^{-2}\,$}}

\def\radpmsq{\hbox{$\rad\m^{-2}\,$}}

\def\kmpspMpc{\hbox{$\kmps\Mpc^{-1}$}}

\ifoldfss
  \ifCUPmtlplainloaded \else
    \NewTextAlphabet{textbfit} {cmbxti10} {}
    \NewTextAlphabet{textbfss} {cmssbx10} {}
    \NewMathAlphabet{mathbfit} {cmbxti10} {} 
    \NewMathAlphabet{mathbfss} {cmssbx10} {} 
  \fi
  \ifAMStwofonts
    \ifCUPmtlplainloaded \else
      \NewSymbolFont{upmath} {eurm10}
      \NewSymbolFont{AMSa} {msam10}
      \NewMathSymbol{\upi}     {0}{upmath}{19}
      \NewMathSymbol{\umu}     {0}{upmath}{16}
      \NewMathSymbol{\upartial}{0}{upmath}{40}
      \NewMathSymbol{\leqslant}{3}{AMSa}{36}
      \NewMathSymbol{\geqslant}{3}{AMSa}{3E}

    \fi
  \fi
\fi 

\ifnfssone
  \newmathalphabet{\mathit}
  \addtoversion{normal}{\mathit}{cmr}{m}{it}
  \addtoversion{bold}{\mathit}{cmr}{bx}{it}
  \newmathalphabet{\mathbfit} 
  \addtoversion{normal}{\mathbfit}{cmr}{bx}{it}
  \addtoversion{bold}{\mathbfit}{cmr}{bx}{it}
  \newmathalphabet{\mathbfss} 
  \addtoversion{normal}{\mathbfss}{cmss}{bx}{n}
  \addtoversion{bold}{\mathbfss}{cmss}{bx}{n}
  \ifAMStwofonts
    \ifCUPmtlplainloaded \else
      \UseAMStwoboldmath
      \makeatletter
      \new@mathgroup\upmath@group
      \define@mathgroup\mv@normal\upmath@group{eur}{m}{n}
      \define@mathgroup\mv@bold\upmath@group{eur}{b}{n}
      \edef\UPM{\hexnumber\upmath@group}
      \new@mathgroup\amsa@group
      \define@mathgroup\mv@normal\amsa@group{msa}{m}{n}
      \define@mathgroup\mv@bold\amsa@group{msa}{m}{n}
      \edef\AMSa{\hexnumber\amsa@group}
      \makeatother
      \mathchardef\upi="0\UPM19
      \mathchardef\umu="0\UPM16
      \mathchardef\upartial="0\UPM40
      \mathchardef\leqslant="3\AMSa36
      \mathchardef\geqslant="3\AMSa3E
    \fi
  \fi
\fi 

\ifnfsstwo
  \DeclareMathAlphabet{\mathbfit}{OT1}{cmr}{bx}{it}
  \SetMathAlphabet\mathbfit{bold}{OT1}{cmr}{bx}{it}
  \DeclareMathAlphabet{\mathbfss}{OT1}{cmss}{bx}{n}
  \SetMathAlphabet\mathbfss{bold}{OT1}{cmss}{bx}{n}
  \ifAMStwofonts
    \ifCUPmtlplainloaded \else
      \DeclareSymbolFont{UPM}{U}{eur}{m}{n}
      \SetSymbolFont{UPM}{bold}{U}{eur}{b}{n}
      \DeclareSymbolFont{AMSa}{U}{msa}{m}{n}
      \DeclareMathSymbol{\upi}{0}{UPM}{"19}
      \DeclareMathSymbol{\umu}{0}{UPM}{"16}
      \DeclareMathSymbol{\upartial}{0}{UPM}{"40}
      \DeclareMathSymbol{\leqslant}{3}{AMSa}{"36}
      \DeclareMathSymbol{\geqslant}{3}{AMSa}{"3E}
    \fi
  \fi
\fi 

\ifCUPmtlplainloaded \else
  \ifAMStwofonts \else 
    \def\upi{\pi}
    \def\umu{\mu}
    \def\upartial{\partial}
  \fi
\fi

\title{ A {\sl Chandra} observation of the distant radio galaxy B2
0902+343: a powerful obscured active galaxy}
\author[A.C.Fabian \etal]
       {A. C. Fabian, C. S. Crawford and K. Iwasawa  \\
        Institute of Astronomy, Madingley Road, Cambridge CB3 0HA}

\date{Submitted 2001: November }

\pubyear{2002}

\begin{document}

\maketitle

\label{firstpage}

\begin{abstract}
We report a {\sl Chandra} observation of the $z=3.395$ radio galaxy
B2~0902+343. The unresolved X-ray source is centred on the active
nucleus. The spectrum is well fit by a flat power law of photon index of
$\Gamma\sim1.1$ with intrinsic absorption of
$8\times10^{22}$\pcmsq, and an intrinsic 2--10\keV\ luminosity of
$3.3\times10^{45}$\ergps. More complex models which allow for a
steeper spectral index cause the column density and intrinsic
luminosity to increase. The data limit any thermal luminosity of the
hot magnetized medium, assumed responsible for high Faraday rotation
measures seen in the radio source, to less than $\sim 10^{45}$\ergps.

\end{abstract}

\begin{keywords}
X-rays: galaxies -- 
galaxies: active --
galaxies: individual: B2 0902+343 -- 

\end{keywords}

\section{Introduction}
0902+343 at $z=3.395$ was the highest redshift radio galaxy known at
the time of its discovery (Lilly 1988). Its radio source structure
consists primarily of a knotted jet extending a short way from a
flat-spectrum nucleus out to the north-west, before being deflected
left through 90\deg\ in a hotspot (Carilli, Owen \& Harris 1994;
Carilli 1995). Further diffuse, very steep-spectrum emission continues
to the north, and two faint hotspots lie symmetrically to the south,
lending the overall structure an inverted-{\sl S} shape. The source is
compact, with the radio emission lying all within 7 arcsec, or
$\sim50$\kpc\ at the redshift of the galaxy (assuming a cosmology of
H$_0$=70\kmpspMpc, $\Omega_0$=1.0 and $\Lambda_0$=0.7). The radio source is
exceptional for its very large Faraday rotation measures (in excess of
1000\radpmsq) and steep rotation measure gradients across the northern
hotspot (Carilli et al 1994; Carilli 1995).

The optical continuum from the host galaxy is visible only from two
clumps to either side of the radio jet (Eisenhardt \& Dickinson 1992),
with extended faint diffuse emission spreading out to the northwest
(Pentericci et al 1999). There is no continuum associated with the
radio core, suggesting that the bulk of the galaxy is totally
obscured. The low optical surface brightness and flat optical-infrared
spectral energy distribution have led authors to propose this to be a
young, still-forming galaxy (Eisenhardt \& Dickinson 1992; Eales et al
1993). The source also has a large luminous Lyman-$\alpha$ halo, also
distributed into two clumps lying to the north of the nucleus, and
bisected by the radio jet (Eisenhardt \& Dickinson 1992). Associated
neutral hydrogen 21cm absorption has been detected against the
(extended) radio
continuum (Uson, Bagri \& Cornwell 1991; Briggs, Sorar \& Taramopoulos
1993; de Bruyn 1996).

0902+343 was detected in a 33~ksec-long exposure with the {\sl ROSAT}
PSPC, but only above 0.7\keV\ (corresponding to $>3$\keV\ in the rest
frame), which could indicate intrinsic absorption of around
$10^{23}$\pcmsq (Crawford 1998). However, the lack of counts in the
{\sl ROSAT} spectrum meant that the low-energy deficit (and thus the
interpretation of the X-ray spectrum as an absorbed power-law) was
significant only at the $\sim2\sigma$ level.

\section{Observation and Results}

\begin{figure}
\psfig{figure=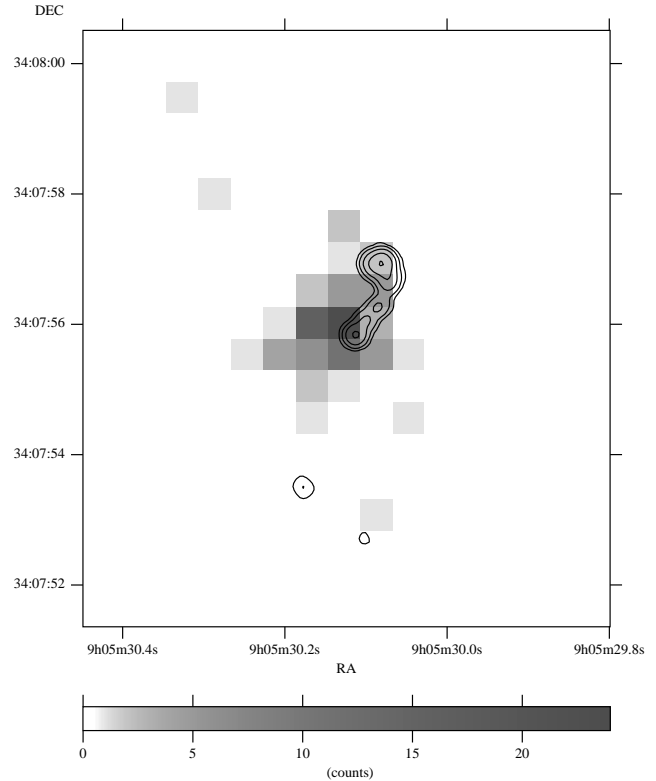,width=0.5\textwidth,angle=0}
\caption{\label{fig:xrayim}
A 0.3-7\keV\ image of the {\sl Chandra} source associated with the
radio galaxy B2 0902+343, with the contours of radio emission at
8\,GHz from Carilli (1995) superposed. The X-ray pixels are 0.5 arcsec
on a side, and the radio contour levels start at a value of
$2\times10^{-4}$ Jy/beam, increasing in steps of a factor of 3.3. }
\end{figure}

0902+343 was observed by the {\sl Chandra} satellite on 2000 Oct 26
for 9.78~ksec with the ACIS-S (sequence number 700212). There were no
background flares during the observation so we extracted data from the
full exposure. An X-ray source is detected at the position of 0902+343
with 94$\pm$10 counts
(Fig~\ref{fig:xrayim}). Background was estimated from a neighbouring,
source-free region. The counts are almost evenly divided between
the 0.5--2\keV\ and 2--7\keV\ energy bands, with 52$\pm7$ and 42$\pm6$
counts respectively. The centroid of the X-ray source (at RA 09 05
30.14, Dec +34 07 56.0) is exactly aligned with the flat spectrum knot
at the southern end of the radio jet (RA 09 05 30.13, Dec +34 07
56.1), supporting its identification as the active nucleus. (We have
confirmed the accuracy of the {\sl Chandra} coordinates by checking
the positions of serendipitous sources in the field of view that are
clearly identified with optical sources; we find the agreement to be
good to within 0.3 arcsec.) The detection is consistent with being a
point source at this position, although we note that a visual
inspection suggests a slight (but not significant) elongation to the
source along the same direction as the radio source axis.

\begin{figure}
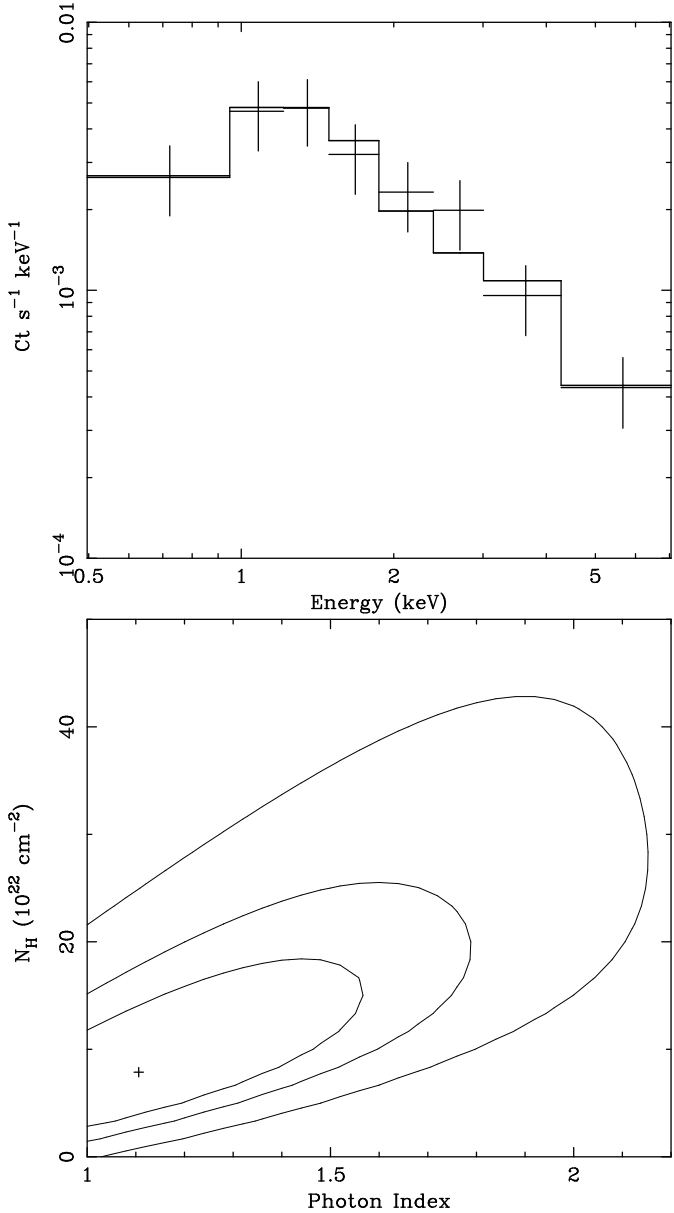

\vbox{
\psfig{figure=spect.ps,width=0.5\textwidth,angle=270}
\psfig{figure=n-g_cont.ps,width=0.5\textwidth,angle=270}}
\caption{\label{fig:xraysp} 
The best-fitting {\sl Chandra} spectrum of the X-ray source associated with the
radio galaxy B2 0902+343, folded through the instrument reponse (top)
and the contours of confidence for the excess absorption N$_{\rm H}$
and photon index $\Gamma$ (lower plot). The contours are for 
$\Delta\chi^2$ = 2.3, 4.61 and 9.21 (corresponding to the
68, 90 and 99 per cent confidence regions for two
interesting parameters). 
The best fit marked by the
cross in the confidence plot is shown against the spectrum as a solid
line. }
\end{figure}

We extract a spectrum for the source, binning it into 12 counts per
bin (Fig~\ref{fig:xraysp}). 
If only Galactic absorption (of column density
2.6$\times10^{20}$\pcmsq; Stark et al 1992) is assumed, then the
spectrum is fit by a power law of improbable photon index of
$\Gamma=0.68\pm0.23$ (uncertainties are 90 per cent for one
interesting parameter). It is much
flatter than the spectrum of most quasars and radio galaxies (eg
Sambruna, Eracleous \& Mushotzky 1999). If we allow for intrinsic
absorption at the redshift of the object we find a good fit for
$\Gamma=1.1^{+0.5}_{-0.4}$ and an inferred intrinsic column density of
$7.8^{+11.8}_{-5.9}\times10^{22}$\pcmsq. The reduced $\chi^{2}$ is
0.363 for five degrees of freedom (acceptable at the 15 per cent
level), and the intrinsic 2--10\keV\ luminosity, (after correction for
absorption) is $3.3\times10^{45}$\ergps. Due to the low number of
photons this spectrum can only be an approximate guide to the
intrinsic properties of the source.  The upper limit to the equivalent
width of any redshifted 6.4\keV\ iron line is 840\eV.

The flat power law of photon index $\Gamma=1.1$ could be due to the
emission being dominated by the jet and similar to that in high
redshift blazars (e.g. Fabian et al 2001a,b). Reeves et al (2001) find
$\Gamma=1.27$ for a radio-loud quasar at $z=3.1$ and attribute it to a
face-on jet. However, Carilli (1995) argues that the jet in 0902 is
pointed out of the Sky plane at an angle of 45-60 degrees so it should
not appear as a blazar.

We have also investigated more complex emission models. Since we are
observing up to about 30~keV in the rest frame of the source,
reflection may be flattening the spectrum. We have therefore refit the
data including a pexrav model (Magdziarz \& Zdziarski 1995). This
yields $\Gamma=1.4^{+0.5}_{-0.4}$ and $N_{\rm H}=9.7^{+12}_{-6}\times
10^{22}\pcmsq$ ($\chi^2/dof=2.3/5$) when the reflection fraction is
assumed to be unity (ie reflection from a flat surface). We have also
examined partial-covering models and find acceptable fits for
$\Gamma=1.4\pm0.7$ with an $\sim80$ per cent covering fraction and
column density of $N_{\rm H}\sim3\times 10^{23}\pcmsq$.

The lack of spatial extent and the good fit of a highly absorbed
power-law to the X-ray spectrum clearly show that the emission that
{\sl Chandra} observes is dominated by the central nucleus. The large
column density reinforces the idea that we are not detecting a very
extended structure, or the mass of absorbing gas would be improbably
large. The very high Faraday rotation measures seen in the radio
source are comparable to those seen locally in cluster cooling flows,
and do suggest the presence of a surrounding hot magnetized medium
(Carilli et al 1994; Carilli 1995). We attempt to constrain the
presence of a thermal component by adding thermal emission to the
absorbed power-law model in the fit to the spectrum. Assuming a
(rest-frame) temperature $kT$=3\keV\ for any thermal component, we
find, assuming $\Gamma=1.8,$ a thermal gas luminosity of
$7^{+3}_{-5}\times 10^{44}\ergps$; it increases to $10^{45}$\ergps if
a higher temperature of $kT$=5\keV\ is assumed. Again the column
density and intrinsic luminosity of the nucleus are increased. The
X-ray data cannot rule out the additional presence of a sizeable mass
of hot, thermal gas, but do not require it. If the central galaxy of a
massive subcluster is forming here then it should be embedded in hot
cooling gas with a temperature of 1-5\keV. Given the high redshift of
the source, clear evidence for a low-energy spectral upturn is
difficult given the declining sensitivity of the ACIS-S detector below
0.5~keV. Further observations with X-ray missions with better
sensitivity at low energies -- such as {\sl XMM-Newton} -- are
required to test for the presence of any extended component of soft
X-ray emission.


\section{Discussion}

The nucleus of 0902+343 is both powerful and highly absorbed. No
object in the sample of Sambruna et al (1999) has a similar
combination of extreme power and column density. Some objects in the
sample of high redshift radio-loud quasars of Cappi et al (1999) do
approach this combination, but the lack of absorption in one of their
absorbed objects, PKS\,0537-286, when observed at much higher
signal-to-noise with XMM-Newton by Reeves et al (2001) raises 
uncertainties. The 2--10\keV\ intrinsic luminosity of 0902+343
is an order of magnitude more than that from the nucleus in Cygnus~A
(Young et al 2002).  If we assume, following the spectral energy
distributions of Elvis et al (1994), that the 2--10~keV luminosity is
about 3 per cent of the total then the bolometric power of the nucleus
is about $10^{47}\ergps$. Much of it is absorbed and presumably
reradiated in the far infrared (FIR) bands. Hughes, Dunlop \& Rawlings
(1997) estimate a value of $10^{13.24}\Lsun$ for the far infrared
luminosity of the source which, after a small correction for the
difference in cosmology used, is $8\times 10^{46}\ergps$ and
comparable with that deduced above. This assumes thermal dust
emission. They have since shown (Archibald et al 2001) that the submm
spectrum is dominated by synchrotron emission, so the above FIR
luminosity should be seen as an upper limit, unless the radiating dust
is warm ($>$100K). Pentericci et al (1999) suggest that the optical
morphology of 0902+343 may be due to obscuration by large amounts of
dust; such dust, and associated gas, may account for the X-ray
absorption.


There is
little room for any strong starburst emission and we conclude that the
active nucleus dominates the luminosity. The central black hole, if
the Eddington limit is not to be exceeded and beaming is unimportant,
should have a mass greater than about $10^9\Msun$. This implies that a
full-grown active nucleus resides at the centre of what has been
inferred to be a proto-galaxy from its appearance at optical and
infrared wavelengths.

\section*{Acknowledgments}

ACF and CSC thank the Royal Society for financial support. We thank
Chris Carilli for providing us with the radio image used in Figure 1,
and Jim Dunlop for information on the SCUBA flux. 

{}

\label{lastpage}

\end{document}